\documentclass[aps,prd,twocolumn,amsmath,amssymb,floatfix,nofootinbib]{revtex4}

\usepackage{graphicx}
\usepackage{dcolumn}
\usepackage{bm}
\usepackage{xcolor}
\usepackage{threeparttable}

\newcommand{\be}{\begin{eqnarray}}
\newcommand{\non}{\nonumber \\}
\newcommand{\ee}{\end{eqnarray}}

\newcommand{\Bell}{\ensuremath{\boldsymbol\ell}}
\newcommand{\BTT}{\ensuremath{ \langle BTT \rangle}}
\newcommand{\BB}{\ensuremath{ \langle BB \rangle}}

\usepackage{hyperref}

\newcommand{\beq}{\begin{equation}}
\newcommand{\eeq}{\end{equation}}
\newcommand{\barr}{\begin{eqnarray}}
\newcommand{\earr}{\end{eqnarray}}
\newcommand{\bs}{\boldsymbol}

\begin{document}

\title{On CMB $B$-Mode Non-Gaussianity}
\author{P. Daniel Meerburg$^{1}$}
\email{meerburg@cita.utoronto.ca}
\author{Joel Meyers$^{1}$}
\email{jmeyers@cita.utoronto.ca}
\author{Alexander van Engelen$^{1}$}
\email{engelen@cita.utoronto.ca}
\author{Yacine Ali-Ha\"{i}moud$^{2}$}
\email{yacine@jhu.edu}
\affiliation{$^1$CITA, University of Toronto, 60 St. George Street, Toronto, Canada}
\affiliation{$^2$Department of Physics and Astronomy, Johns Hopkins University, Baltimore, MD 21218, USA}

\begin{abstract}
  We study the degree to which the cosmic microwave background (CMB) can be used to constrain primordial non-Gaussianity involving one tensor and two scalar fluctuations, focusing on the correlation of one polarization $B$ mode with two temperature modes.  In the simplest models of inflation, the tensor-scalar-scalar primordial bispectrum is non-vanishing and is of the same order in slow-roll parameters as the scalar-scalar-scalar bispectrum.  We calculate the \BTT\ correlation arising from a primordial tensor-scalar-scalar bispectrum, and show that constraints from an experiment like CMB-Stage IV using this observable are more than an order of magnitude better than those on the same primordial coupling obtained from temperature measurements alone.  
We argue that $B$-mode non-Gaussianity opens up an as-yet-unexplored window into the early Universe, demonstrating that significant information on primordial physics remains to be harvested from CMB anisotropies.  
\end{abstract}

\maketitle

\section{Introduction}\label{sec:Introduction}

Observations of the cosmic microwave background (CMB) and large-scale structure have in recent years greatly advanced our understanding of the contents and history of the Universe. Current observational data fit well with the concordance six-parameter $\Lambda$CDM model \cite{Ade:2015xua}. While the excellent agreement between the model and the data is undoubtedly a triumph of modern cosmology, our understanding of the Universe's initial conditions remains limited, and it is crucial to explore new observational probes that can deepen our understanding of the underlying physics. 


Within the concordance model, the initial fluctuations are fully accounted for by primordial density fluctuations, which are purely Gaussian, adiabatic, and nearly scale invariant, and can be described by just two parameters, the amplitude $A_s$ and the scalar spectral index $n_s$.  A very wide class of early-universe models are capable of accounting for such primordial fluctuations, and therefore these two parameters alone are not greatly informative. A great deal of effort has therefore been devoted to searching for signatures of deviations from this simple picture.  Even in the absence of a detected deviation from the concordance model, upper limits on various observables greatly help to discriminate among early-universe models.  Particularly interesting observables in this regard include non-adiabaticity \cite{PlanckInflation2015}, running of the scalar spectral index \cite{RunningSpectralIndex,RunningSpectralIndex2,RunningSpectralIndex3,Pajer2016}, primordial tensor fluctuations \cite{Kamionkowski:1996zd,Seljak:1996gy,Zaldarriaga:1996xe,Kamionkowski:1996ks}, and non-Gaussianity \cite{KomatsuNGs,PlanckNonGaussianity2015,LSSnonGaussianity2014}.
\begin{table*}[t]\label{table:Parity}
  \begin{center}
    \begin{tabular}{ | c || c | c |}
      \hline
      Full-sky & $\sum_n\ell_n =$ even & $\sum_n\ell_n =$ odd \\ \hline
      Flat-sky & left-handed $=$ right-handed & left-handed $= (-)$ right-handed \\ \hline \hline
      Non-vanishing  & $\langle TTT\rangle$, $\langle TEE\rangle$, $\langle TTE \rangle$,  & $\langle BTT\rangle$, $\langle BEE\rangle$,   \\ 
      in parity-conserving universe & $\langle EEE \rangle$, $\langle BBE \rangle$, $\langle BBT \rangle$ & $\langle BET \rangle$, $\langle BBB \rangle$ \\ \hline
    \end{tabular}
  \end{center}
  \caption{Properties of full-sky and flat-sky three-point functions in a parity-conserving universe. The first column contains three-point functions studied in the standard analysis.  The second column (and in particular $\langle BTT \rangle$) is the focus of this work. There are additional non-vanishing three-point functions when parity conservation is violated which are studied e.g. in Refs.~\cite{ParityOddBispectra1,ParityOddBispectra2}. }
  \label{default}
\end{table*}

The CMB contains cosmological information both in its temperature and linear polarization. The polarization field can be separated into $E$-modes and $B$-modes which have opposite intrinsic parity.  Primordial scalar fluctuations source temperature fluctuations and $E$-mode polarization, while primordial tensors source $T$, $E$, and $B$ fluctuations \cite{Kamionkowski:1996zd,Seljak:1996gy,Zaldarriaga:1996xe,Kamionkowski:1996ks}.

The CMB temperature power spectrum has recently been measured to cosmic variance limits up to multipole $\ell \approx 2000$ using data from the Planck satellite \cite{PlanckInflation2015}. Due to the diffusion damping of fluctuations on smaller scales, we do not expect that lower-noise observations of the CMB will provide significantly more cosmological information from temperature fluctuations.  There is some additional information which can be gained from lower-noise measurements of $E$-mode polarization of the CMB, though temperature and $E$-mode fluctuations are sourced by nearly the same cosmological modes, and therefore constraints on observables like non-Gaussianity and the running of the spectral index will not significantly improve on the current status with CMB measurements alone \cite{Komatsu:2002db}. Alternatively, large-scale-structure observations can provide additional cosmological information which could eventually lead to a detection of non-Gaussianity or running \cite{LSSnonGaussianity2014}. This will require, however, overcoming significant challenges in the modeling of non-linearities \cite{PeacockDodds1996,CooraySheth,LSSnonGaussianity2014}, biasing \cite{ShethBias,TinkerBias}, and complex astrophysical processes \cite{LymanAlphaCsomology,AstroGalaxies, LargeRelSims1,LargeRelSims2,SchayeLargeRelSims}. Another potential avenue for measurements of primordial non-Gaussianities is the tomographic mapping of neutral hydrogen at high redshift with the 21-cm line \cite{Cooray_06, Pillepich_07, 21cmBispectrum}. This will require overcoming daunting observational challenges \cite{21cmCosmology, Fur2009proc,Mao:2008ug}, as well as a detailed modeling of the intrinsic non-linearities of 21-cm fluctuations \citep{21cmBispectrum}. Finally, spatial fluctuations of CMB spectral distortions can also be used to probe primordial non-Gaussianities \cite{Pajer_12, Emami_15, Khatri_15}.

Until recently, the best constraint on primordial tensor fluctuations was derived from the measurement of the CMB temperature power spectrum \cite{WMAP2013, PlanckInflation2013}.Unfortunately, on large angular scales where the tensor contribution to the temperature power spectrum is most significant, the scalar contribution to the temperature power spectrum is much larger, and constraints on tensor fluctuations from temperature measurements alone are hindered by the relatively large cosmic variance of the temperature fluctuations.  

Ongoing and future observations of the CMB will drastically improve the constraints on primordial tensor fluctuations by searching for $B$-mode polarization on large angular scales.
Several such experiments are currently underway, with the most recent constraints coming from the BICEP/Keck experiment \cite{Bicep2015}.  The biggest astrophysical obstacles in constraining  the primordial signal are the contributions from dust \cite{FlaugerDust,PlanckDust2014} and lensing of $E$-modes to $B$-modes \cite{seljak2004}. For the former, we will have to rely on multi-frequency information to separate the dust component from the primordial signal.  For the  latter, delensing will become crucial for removing lens-induced fluctuations and requires a high-fidelity lensing map \cite{Seljak:2003pn}.

While the usual searches for non-Gaussianity focus on the $N$-point statistics of scalar fluctuations, in this paper we will discuss the relatively unexplored observational signatures of non-Gaussian correlations involving tensor fluctuations.  Since tensor fluctuations source $T$, $E$, and $B$ fluctuations, observational searches for bispectra constructed from $T$ and $E$ fluctuations naturally place constraints on both scalar and tensor non-Gaussianity.  Just as in the case of the power spectrum, the contributions to $T$ and $E$ fluctuations from scalars are much larger than those of tensors, and so constraints on tensor non-Gaussianity with these bispectra are relatively weak.  On the other hand, bispectra involving primordial $B$-mode fluctuations are sourced by tensor non-Gaussianity but not by scalar non-Gaussianity, and are therefore capable of providing a much tighter constraint on tensor non-Gaussianity.  Since observations of $B$-modes are not presently cosmic variance-limited, there is a great deal of room for improvement with future observations of the CMB polarization.  This reasoning strongly motivates searching for bispectra involving primordial $B$-modes as a probe of primordial non-Gaussianity.  In this paper, we explore in detail how the $\langle BTT \rangle$ bispectrum can be used to constrain tensor non-Gaussianity and thereby give us insight into the physics of the early universe.

The primordial tensor-scalar-scalar bispectrum is naturally non-vanishing, and in fact is of the same order in slow-roll parameters as the primordial scalar-scalar-scalar bispectrum in the simplest models of single-field slow-roll inflation \cite{Maldacena2002}. In more general models, the shape and amplitude of the tensor-scalar-scalar bispectrum can differ quite significantly from those predicted in the simplest models, so observational constraints on this quantity give non-trivial insight into the physics of the early Universe \cite{Gao:2012ib}.  The primordial tensor-tensor-scalar and tensor-tensor-tensor bispectra are also non-vanishing in single-field slow-roll inflation, as well as in more general models \cite{Maldacena2002,Gao:2012ib,Maldacena:2011nz}.  Primordial non-Gaussianity involving tensors provides a set of observables which are distinct from and complementary to scalar non-Gaussianity.  Also, just as in the case of scalars \cite{Komatsu:2002db}, there is in principle much more information in tensor non-Gaussianity than in the tensor power spectrum alone.

Despite the differing intrinsic parity of temperature fluctuations and $B$-modes, the $\langle BTT \rangle$ bispectrum is non-vanishing for particular combinations of multipoles.  To be more specific, under spatial inversion the multipole coefficients for $T$, $E$, and $B$ transform as \cite{weinberg2008cosmology}
\begin{align}\label{ParityProperties}
  a^T_{\ell m}& \rightarrow (-1)^\ell a^T_{\ell m} \, , \non
  a^E_{\ell m}& \rightarrow (-1)^\ell a^E_{\ell m} \, , \non
  a^B_{\ell m}& \rightarrow (-1)^{\ell+1} a^B_{\ell m}. \nonumber
\end{align}
These properties along with statistical isotropy imply that $\langle a^B_{\ell m}a^T_{\ell' m'}\rangle = \langle a^B_{\ell m}a^E_{\ell' m'}\rangle = 0$ in a parity-conserving universe since these quantities change sign under spatial inversion.  On the other hand, we find that under spatial inversion, the bispectrum of interest transforms as
\be
\label{BispectrumParity}
\langle a^T_{\ell_1 m_1}a^T_{\ell_2 m_2}a^B_{\ell_3 m_3}\rangle &\rightarrow& (-1)^{\ell_1+\ell_2+\ell_3+1} \langle a^T_{\ell_1 m_1}a^T_{\ell_2 m_2}a^B_{\ell_3 m_3}\rangle, \nonumber
\ee
which therefore must vanish in a parity-conserving universe for $\sum_n\ell_n=\,$even but not for $\sum_n\ell_n=\,$odd (see Table~\ref{table:Parity}).

The above remarks straightforwardly generalize to all forms of non-Gaussianity.  In a parity-conserving and statistically isotropic universe, any connected $N$-point function constructed from $T$, $E$, and $B$ fluctuations containing an odd number of $B$-mode fluctuations vanishes for $\sum_n\ell_n=\,$even but not for $\sum_n\ell_n=\,$odd, while those containing an even number of $B$-mode fluctuations vanish for $\sum_n\ell_n=\,$odd but not for $\sum_n\ell_n=\,$even.  The case of $N=2$ is special since statistical isotropy always implies that $\ell_1+\ell_2=\,$even for two-point statistics.

\begin{figure}[h]
  \includegraphics[width = 85 mm]{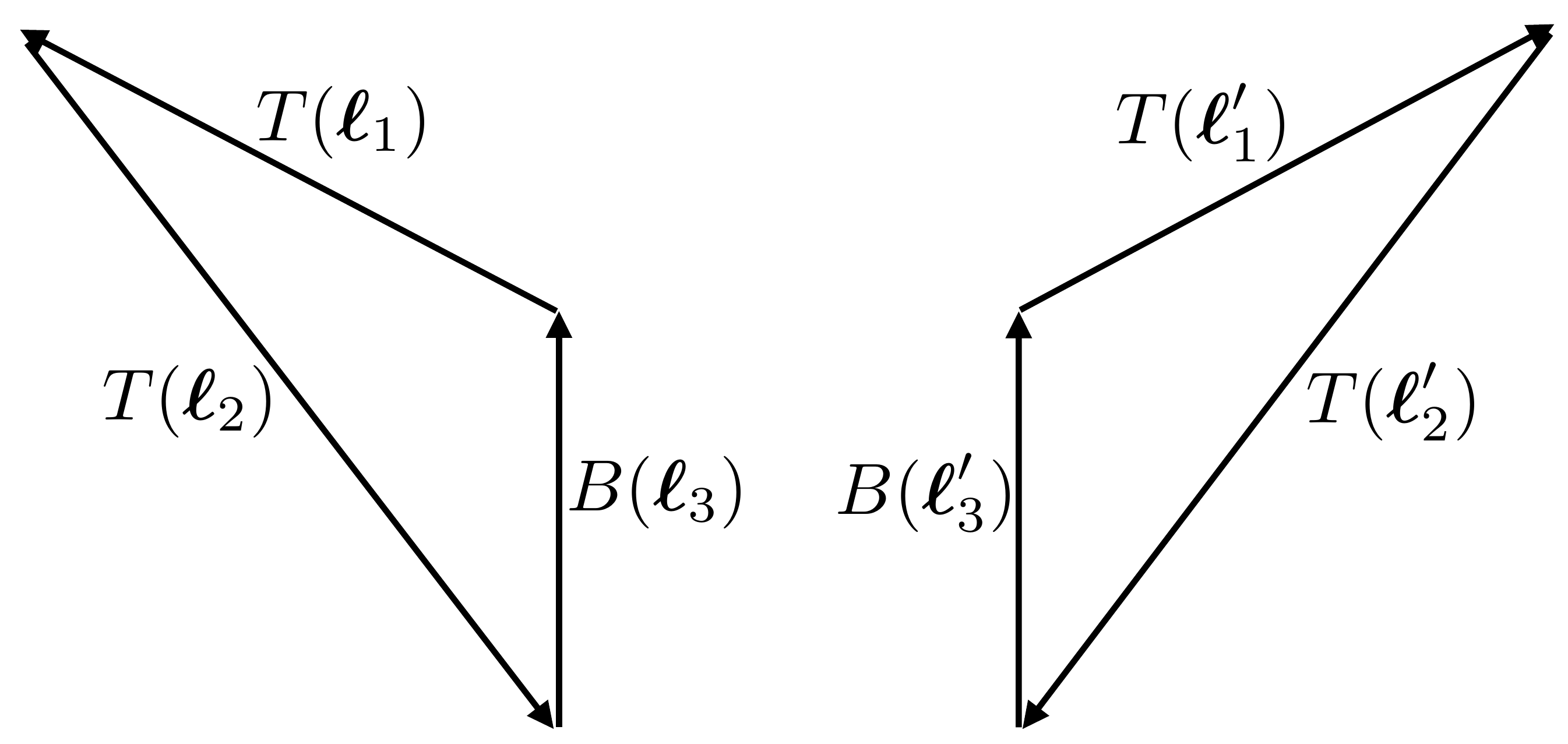}
  \caption{Two triangles in multipole space that are mirrored images of one another. The three-point function $\langle ETT \rangle $ takes the same value on both configuration, whereas $\langle BTT \rangle $ changes sign.}\label{fig:BTT}
\end{figure}

Let us briefly summarize our motivations.  Non-Gaussian CMB statistics involving $B$-mode fluctuations are non-vanishing under standard assumptions about the properties of our Universe. Existing data can be used to place new constraints on these quantities. Present measurements of $B$-modes are not cosmic-variance limited, so upcoming lower-noise CMB polarization data will drastically improve upon our current capabilities in this regard.  Measurements of these statistics can provide non-trivial constraints on primordial non-Gaussianity involving tensor fluctuations.  Primordial tensor non-Gaussianity is in general independent of primordial scalar non-Gaussianity and is therefore a complementary probe of early-universe physics.  

The goal of this paper is to explore the potential of the $\langle BTT \rangle$ bispectrum as a probe of the primordial Universe. In Sec.~\ref{sec:envelope} we discuss the geometric properties of the $\langle BTT \rangle$ bispectrum. In Sec.~\ref{sec:Theory} we review the predictions of single-field slow-roll inflation for the primordial tensor-scalar-scalar bispectrum and discuss its properties. We construct the $\langle BTT\rangle$ bispectrum in the flat-sky limit in Sec.~\ref{sec:BTTflatsky}. We then forecast constraints this observable in Sec.~\ref{sec:SN} for current and future experiments. We discuss the implications and future extensions of our work in Sec.~\ref{sec:Conclusions}.

\begin{figure*}[t] 
  \centering
  \includegraphics[scale=0.31]{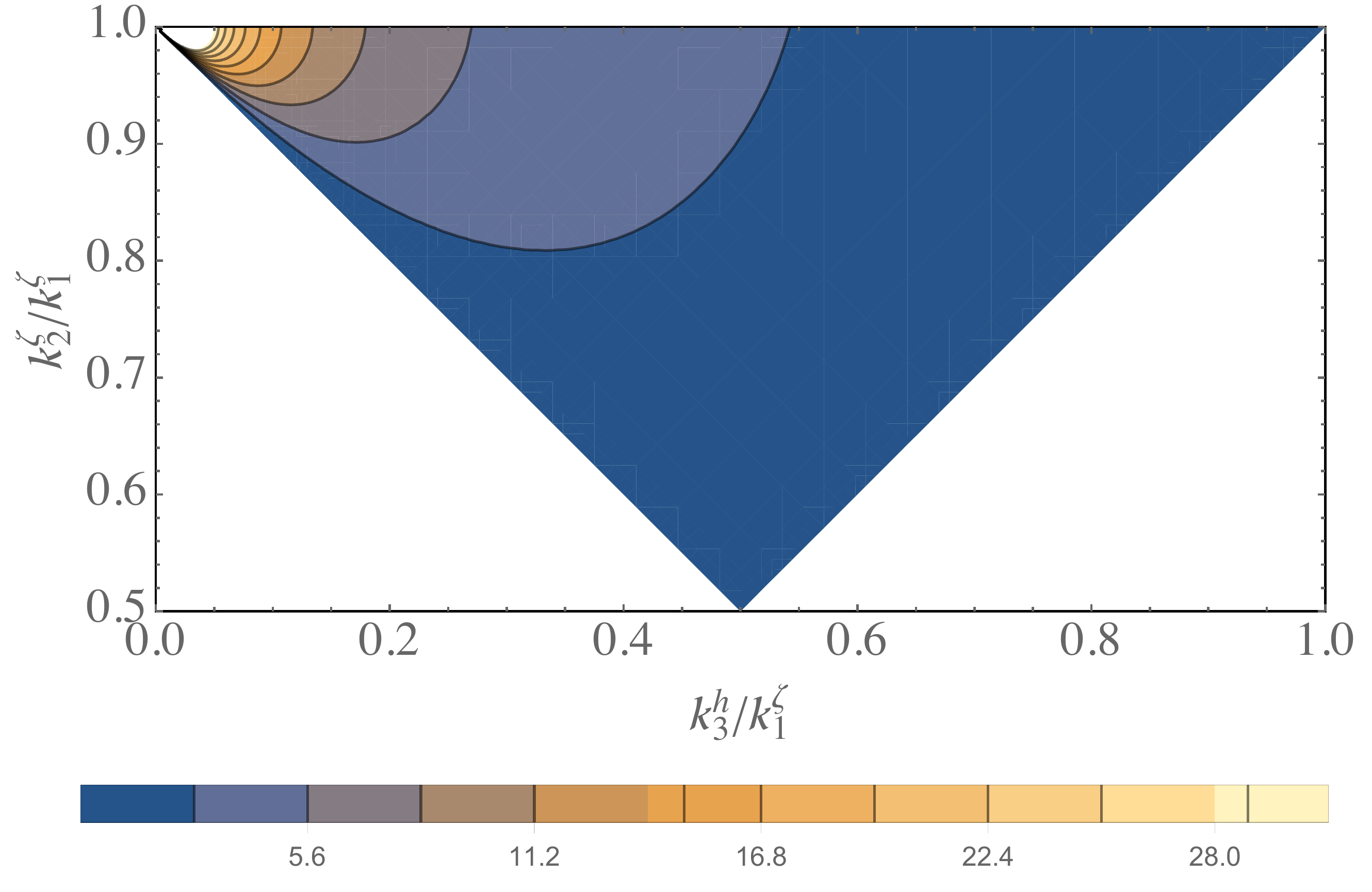} 
  \includegraphics[scale=0.31]{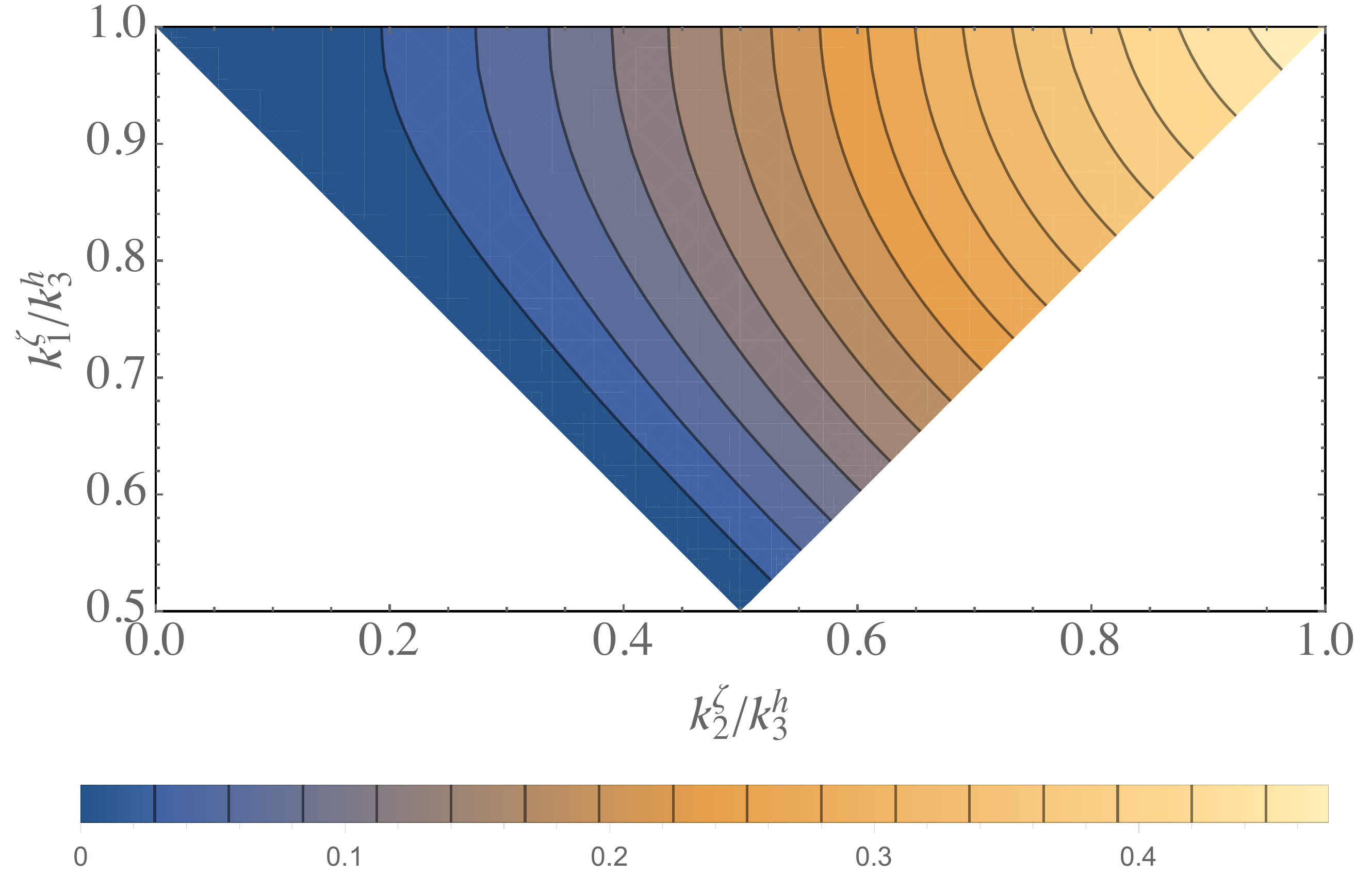} 

   \caption{Two unique slices showing the primordial tensor-scalar-scalar bispectrum from single-field slow-roll inflation with $k_3$ representing the wave vector of the tensor.   Left: The bispectrum is enhanced when $k_3 \ll k_1 \sim k_2$ (top left corner). The enfolded limit, i.e. $k_1 + k_2 = k_3$ (bottom left edge), is suppressed, since then all momenta are aligned. Right: When one or both scalar momenta $k_1$ and $k_2$ are aligned with the tensor momentum $k_3$ (bottom left edge) the spectrum is suppressed as compared to the equilateral configuration (top right).}
   \label{fig:TensorScalarBispectrum}
\end{figure*}

\section{Geometric properties of the $\langle BTT \rangle$ bispectrum}\label{sec:envelope}

In this Section we outline the general geometric properties of the $\langle BTT \rangle$ correlation function in the flat-sky approximation. We start by considering properties of correlation functions involving the polarization tensor $P^{ab}$.

\subsection{Correlation functions involving the polarization tensor $P^{ab}$}

Let us first consider the correlation function $\langle T(\bs{x}_1) P^{ab}(\bs{x}_2) \rangle$. Statistical homogeneity implies that this can only be a function of $\bs{x}_{12} \equiv \bs{x}_1 - \bs{x}_2$. Since $P^{ab}$ is a symmetric, trace-free tensor, so must be this correlation function, which must therefore take the form
\beq
\langle T(\bs{x}_1) P^{ab}(\bs{x}_2) \rangle = F(x_{12}) \left[2 \hat{x}_{12}^a \hat{x}_{12}^b - \delta^{ab} \right],
\eeq
where $F$ only depends on the magnitude of $\bs{x}_{12}$ by statistical isotropy. Using this result, we can easily show that the correlation function in multipole space is of the form
\beq
\langle T(\bs{\ell}) P^{ab}(\bs{\ell}') \rangle = \delta^{(2)}(\bs{\ell}  +\bs{\ell}') G(\ell) \left[2 \hat{\ell}^a \hat{\ell}^b - \delta^{ab} \right]. \label{eq:TPab}
\eeq
Similarly, one can show that statistical homogeneity and isotropy imply that the three-point function $\langle T(\bs{\ell}_1) T(\bs{\ell}_2) P^{ab}(\bs{\ell}_3)\rangle$ takes the form
\barr
&&\langle T(\bs{\ell}_1) T(\bs{\ell}_2) P^{ab}(\bs{\ell}_3)\rangle =  \delta^{(2)}(\bs{\ell}_1 + \bs{\ell}_2 + \bs{\ell}_3) \nonumber\\
&& \times \sum_{i\leq j = 1}^3 G_{ij}(\ell_1, \ell_2, \ell_3) \left[\hat{\ell}_i^a \hat{\ell}_j^b + \hat{\ell}_j^a \hat{\ell}_i^b  - (\hat{\ell}_i \cdot \hat{\ell}_j) \delta^{ab} \right], \label{eq:TTPab}
\earr
where we also used the fact that $P^{ab}$ is symmetric and trace-free. The symmetry of the three-point function under interchange of $\bs{\ell}_1$ and $\bs{\ell}_2$ moreover imposes $G_{22}(\ell_1, \ell_2, \ell_3) = G_{11}(\ell_2, \ell_1, \ell_3)$ and $G_{23}(\ell_1, \ell_2, \ell_3) = G_{13}(\ell_2, \ell_1, \ell_3)$. 

We emphasize that an implicit underlying assumption to derive these general results is that the physics governing temperature and polarization fluctuations is parity-conserving, both at the level of initial conditions, and for their subsequent evolution. Explicitly, this implies that the antisymmetric Levi-Civita tensor $\bs{\epsilon}$ cannot appear in any of the above correlation functions.

\subsection{Implications for $E$ and $B$-modes}

We recall that in the flat-sky limit, the $E$- and $B$-mode decomposition of the polarization tensor field $P^{ab}$ is obtained as follows \cite{KamionkowskiKovetz2015}
\barr
\nabla^2 E &\equiv& \partial_a \partial_b P^{ab},\\
\nabla^2 B &\equiv& \epsilon_{a c} \partial^c \partial_b P^{ab},
\earr
where repeated indices are summed. While $E$ is a scalar quantity, $B$ is a pseudo-scalar, or parity-odd observable: its sign depends on the chosen handedness of the coordinate system. In multipole space, we have
\barr
E(\bs{\ell}) &\equiv& \hat{\ell}_a \hat{\ell}_b P^{ab}(\bs{\ell}), \label{eq:E-flat}\\
B(\bs{\ell}) &\equiv& \epsilon_{ac} \hat{\ell}^c \hat{\ell}_b P^{ab}(\bs{\ell}). \label{eq:B-flat}
\earr
Using these equations into Eq.~\eqref{eq:TPab}, we obtain
\barr
\langle T(\bs{\ell}) E(\bs{\ell}') \rangle &=& \delta^{(2)}(\bs{\ell} + \bs{\ell}') G(\ell),\\
\langle T(\bs{\ell}) B(\bs{\ell}') \rangle &=& 0,
\earr
where the vanishing of the $\langle B T \rangle$ correlation results from the antisymmetry of $\bs{\epsilon}$.

Substituting Eq.~\eqref{eq:E-flat} into Eq.~\eqref{eq:TTPab}, we see that $\langle T(\bs{\ell}_1) T(\bs{\ell}_2) E(\bs{\ell}_3) \rangle$ only depends on the magnitudes $\ell_i$ and on the scalar products between the three multipoles. The triangle condition $\bs{\ell}_1 + \bs{\ell}_2 + \bs{\ell}_3 = \bs{0}$ implies that the scalar product of any two multipoles can be rewritten as a function of the magnitudes only. We therefore obtain 
\beq
\langle T(\bs{\ell}_1) T(\bs{\ell}_2) E(\bs{\ell}_3) \rangle' =  \mathcal{G}(\ell_1, \ell_2, \ell_3),
\eeq
where the prime indicates that we divided by $\delta^{(2)}(\sum \bs{\ell}_i)$ and $\mathcal{G}$ is some function of the magnitudes $\ell_i$, symmetric under exchange of $\ell_1, \ell_2$.

Substituting Eq.~\eqref{eq:B-flat} into Eq.~\eqref{eq:TTPab}, we see that some of the terms vanish (e.g. the $G_{33}$ term), but not all of them. The final form of the $\langle BTT \rangle$ correlation function is 
\barr\label{BTTformH}
\langle T(\bs{\ell}_1) T(\bs{\ell}_2) B(\bs{\ell}_3) \rangle' &=& (\hat{\ell}_1 \times \hat{\ell}_3) ~\mathcal{H}(\ell_1, \ell_2, \ell_3)  \nonumber\\
&+& (\hat{\ell}_2 \times \hat{\ell}_3) ~\mathcal{H}(\ell_2, \ell_1, \ell_3),
\earr
where $\hat{\ell}_1 \times \hat{\ell}_3 \equiv \epsilon_{ac} \hat{\ell}_1^a \hat{\ell}_3^c$. 

Geometrically, this can be understood as follows. Consider two triangles in multipole space that are mirrored images of one another in the way shown in Fig.~\ref{fig:BTT}. The magnitudes of all wavenumbers are identical (hence the scalar products between any two of them). However, the cross products $\hat{\ell}_1 \times \hat{\ell}_3$ and $\hat{\ell}_2 \times \hat{\ell}_3$ take on opposite signs on the two triangles. This implies that the values of the tree-point function $\langle E TT \rangle$ is identical on both triangles, whereas $\langle B TT\rangle$ has the same absolute value but takes a different sign on the two mirrored triangles. A violation of these symmetry properties would be an indicator for parity-violating physics \cite{KamionkowskiSouradeep2010}.

Finally, we note an analogy with gravitational lensing reconstruction.  One could use a cross product of the CMB temperature gradient with itself,  $(\nabla T) \times (\nabla T)$,  to reconstruct the ``curl'' portion of deflection of CMB photons that would arise, e.g., from tensor fluctuations out to the recombination surface  \cite{cooray2005}.   This map would be an estimate of a field with  odd parity, and could be cross-correlated with a map of $B$-mode polarization, which also has odd parity, giving a nonzero result for a universe containing primordial tensors.

\section{Primordial Bispectrum}\label{sec:Theory}
In this section, we will discuss the properties of the primordial tensor-scalar-scalar bispectrum predicted in single-field slow-roll inflation, which were first obtained in Ref.~\cite{Maldacena2002}.  This model provides an example of tensor non-Gaussianity in the simplest of inflationary models, which we will use to motivate a template to compute the $\langle BTT\rangle$ bispectrum. For more general models, the primordial bispectrum will differ from the results presented here, and modifications should be made to the template to address these changes. 

Following the work by Maldacena \cite{Maldacena2002}, for inflation driven by a single slowly-rolling scalar field, the primordial bispectrum of two scalar fluctuations and one tensor fluctuation is given by 
\be
\left\langle \zeta(\bs{k}_1)\zeta(\bs{k}_2)h^{\pm}(\bs{k}_3) \right\rangle &=& (2\pi)^3 F^{00\pm2}(\bs{k}_1,\bs{k}_2,\bs{k}_3) \nonumber \\
&& \times \delta^{(3)} \left(\bs{k}_1 + \bs{k}_2 + \bs{k}_3 \right),
\ee
where we have defined
\be
F^{00\pm2}(\bs{k}_1,\bs{k}_2,\bs{k}_3) \equiv  \frac{H_*^4}{4M_{\mathrm{pl}}^4\epsilon_*} I(k_1,k_2,k_3)e_{ab}^{\mp}(\bs{k}_3)k_1^ak_2^b, ~
\label{eq:FTensScal}
\ee
where $H_*$ and $\epsilon_*$ are, respectively, the Hubble rate and first slow-roll parameter during inflation, and 
\be\label{eq:NGshape}
	 && I(k_1,k_2,k_3) \equiv \frac{1}{k_1^3k_2^3k_3^3} \non
	 && \quad \times \left( -k_t + \frac{k_1k_2+k_2k_3+k_1k_3}{k_t} + \frac{k_1k_2k_3}{k_t^2} \right) \, .
\ee

Let us now examine the transverse traceless polarization tensor $e_{ab}^{\pm}$ more closely.  It is defined such that $e_{ab}^{\lambda}(\hat{k})e_{ab}^{\lambda'}(-\hat{k}) = 2\delta_{\lambda\lambda'}$, and when $\hat{k}$ points in the $z$-direction, it is given by
\be
	e_{ab}^{\pm}(\hat{z}) = \frac{1}{\sqrt2}
	\begin{pmatrix}
   1       & \pm i & 0  \\
   \pm i       & -1 & 0 \\
   0       & 0 & 0 
  \end{pmatrix} \, .\label{eq:e_ab}
\ee

We write $\hat{k}_i = (\sin \Theta_i \cos \phi_i,\sin \Theta_i \sin \phi_i,\cos \Theta_i)$ for $i = 1, 2$, with $0\leq \Theta_i\leq \pi$ and $0\leq \phi_i\leq 2\pi$. The triangle constraint $\bs{k}_1 + \bs{k}_2 + \bs{k}_3 = 0$ imposes $\phi_1 = \phi_2 \equiv \phi$, while $\Theta_1$ and $\Theta_2$ are related through
\be
\sin \Theta_2 = -\frac{k_1}{k_2} \sin \Theta_1.
\ee
We then find 
\be
e^{(\mp)}_{ab}(\bs{k}) \hat{k}_1^a \hat{k}_2^b &=& -\frac{\sin \Theta_1 \sin \Theta_2}{\sqrt{2}} e^{\mp 2 i \phi} \nonumber \\
&=& -\frac{k_1}{k_2} \frac{(\sin \Theta_1)^2}{\sqrt{2}} e^{\mp 2 i \phi}.
\ee
We show $F^{00+2}(\bs{k}_1,\bs{k}_2,\bs{k}_3) k_1^2k_2^2k_3^2$ for $\phi = 0$ in Fig.~\ref{fig:TensorScalarBispectrum}. The tensor-scalar-scalar bispectrum has two different squeezed limits: one in which the wavenumber of the tensor is much smaller than those of the scalars ($k_h \ll k_{\zeta_1}\sim k_{\zeta_2}$), and another in which the wavenumber of one of the two scalars is much smaller than those of the other scalar and of the tensor ($k_{\zeta_2} \ll k_h \sim k_{\zeta_1}$). In the former case, if the scalar wavenumbers are perpendicular to the tensor wavenumber the bispectrum is enhanced. Conversely, the enfolded limit, when both scalar wavenumbers are roughly aligned with the tensor wavenumber the bispectrum is suppressed by the polarization sum.  

We propose a reference definition of the primordial tensor-scalar-scalar bispectrum of the form 
\be
\left\langle \zeta(\bs{k}_1)\zeta(\bs{k}_2)h^{\pm}(\bs{k}_3) \right\rangle &=& (2\pi)^3 16 \pi^4 A_s^2 \sqrt{r}f_\mathrm{NL}^{h\zeta\zeta}  \delta^{(3)} \left(\sum_{n=1}^3\bs{k}_n\right) \non
&& \times\mathcal{I}(k_1,k_2,k_3) e_{ab}^{\mp}(\bs{k}_3)\bs{k}_1^a \bs{k}_2^b, 
\label{eq:hzz_generalform}
\ee
with $ \mathcal{I}(k,k,k) \propto k^{-8}$ in a scale-invariant universe. For single-field slow-roll inflation,  $ \mathcal{I}(k_1,k_2,k_3) = I(k_1,k_2,k_3)$ and $ f_\mathrm{NL}^{h\zeta\zeta}  = \sqrt{r}/16 $, but these quantities will differ in more general models.

In the squeezed limit where the tensor wavenumber is much smaller than the wavenumbers of the scalars, the properties of the bispectrum in single-field slow-roll inflation are entirely determined by the fact that the long-wavelength tensor fluctuation is locally equivalent to an anisotropic rescaling of coordinates \cite{Maldacena2002}.  Similar to the case of scalar non-Gaussianity, this implies that there exists no locally observable mode coupling between long-wavelength tensors and short-wavelength scalar fluctuations in single-field inflation \cite{Creminelli:2004pv, Pajer:2013ana}.  This tensor consistency condition applies more broadly than the more familiar scalar consistency condition, since the same logic will apply to any scalar field minimally coupled to gravity, whether or not its energy density drives inflation \cite{Gao:2012ib}.  For this reason, if single-field slow-roll inflation or something similar is responsible for the primordial fluctuations we observe, we are unlikely to gain much insight into the physics from the squeezed limit of the tensor-scalar-scalar bispectrum, but it remains interesting to search for deviations from the predictions of the simplest models.  Despite these subtleties regarding the squeezed limit, in what follows we will take Eq.~(\ref{eq:NGshape}) as our primordial template.

\subsection{On the normalization of $f^{h \zeta\zeta}_{\rm NL}$}\label{sec:norm}

In the above we define $f^{h \zeta\zeta}_{\rm NL} \sim \langle h\zeta\zeta\rangle /\langle \zeta \zeta\rangle^{3/2} \langle hh\rangle^{1/2}$. Other choices can be found in the literature, but can all be related to one another via a simple calculation. Our reasoning for the definition above is two-fold. First, from a primordial perspective, naively we would expect the amplitude of the non-Gaussian signal to be proportional to $\sqrt{r}$ given the presence of a single $h$. Secondly, in this way the measured bispectrum and it amplitude will behave similar to the amplitude of the tensor power spectrum when measured in the cosmic variance limit. Using $\BTT$ as our measure, the variance $\sigma_{BTT} \propto \sqrt{r}$ in this limit. Therefore $\sigma(f_{\rm NL}^{h\zeta\zeta})$ will be constant in the cosmic variance limit and only change as a function of $\ell_{\rm max}$, the maximum number of observed modes on the sky.


 \begin{figure*}[t] 
  \centering
  \includegraphics[scale=0.35]{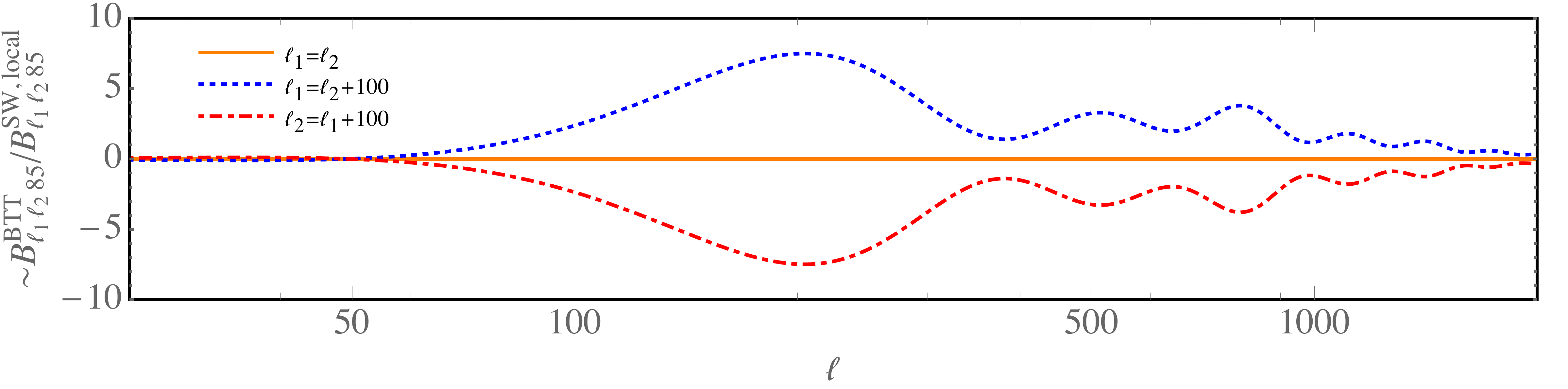}    
  \caption{The $\langle BTT \rangle$ bispectrum in the fly-sky limit computed from Eq.~\eqref{eq:BTTbispectrum2} normalized using the analytical form of the local $\langle TTT \rangle$ bispectrum.  We see explicitly that the $\langle BTT \rangle$ bispectrum changes sign under the interchange of of the two $T$ multipoles $\ell_1$ and $\ell_2$ and vanishes when $\ell_1 = \ell_2$. 
The overall amplitude of $\langle BTT \rangle$ for our chosen template is almost three orders of magnitude smaller than the local template $\langle TTT \rangle$ bispectrum from scalars.}
  \label{fig:BTTbispectrum}
\end{figure*}

\section{$\langle BTT \rangle$ in the flat-sky limit}\label{sec:BTTflatsky}

The full-sky CMB $\langle BTT \rangle$ bispectrum has been worked out before in Ref.~\cite{MaresukeThesis,Maresuke2010,Maresuke2011} and using total angular momentum spherical harmonics in Ref.~\cite{TotalAnhularMomentumWaves2012}. However, both results lead to expressions that are not very transparent regarding the symmetries of this correlation function. 
Here instead, we derive the spectrum in the flat-sky approximation were these symmetries, which we derived in Sec.~\ref{sec:envelope}, are immediately apparent. The results presented here are accurate as long as we restrict ourselves to scales $\ell \gtrsim 10$. For the computation of the signal-to-noise ratio in the next section, we will use this approximate form, which should provide reasonably accurate results since our forecasts focus on experiments that will most likely not be able to map out the lowest multipoles. The flat-sky definition of the bispectrum takes the form \cite{NonGaussianShapes}
\be
\langle a_T(\Bell_1) a_T(\Bell_2)  a_B(\Bell_3) \rangle =(2\pi)^2 \delta^{(2)}(\sum \Bell_i)B^{TTB}_{\Bell_1 \Bell_2 \Bell_3}. 
\label{eq:flatsky}
\ee

In the flat-sky approximation the temperature fluctuation arising from primordial scalar perturbations is given by \cite{MaresukeThesis,Maresuke2010,Maresuke2011}
\be
a_T^\zeta(\Bell) &=& \int_{0}^{\tau_0}\mathrm{d}\tau \int \frac{\mathrm{d}^3k}{(2\pi)^3} \zeta(\bs{k}) e^{-i k^z D} \nonumber \\
&& \times S_{T}^{\zeta}(\bs{k},\tau) (2\pi)^2\delta^{(2)}\left(\bs{k}^\parallel D-\Bell\right) \, ,
\label{eq:al_T}
\ee
where $\tau$ is the conformal time, $\tau_0$ is its value today, $D \equiv \tau_0 - \tau$, $\bs{k}^\parallel$ is the component of $\bs{k}$ parallel to the plane of the sky, $k^z$ its projection along the line of sight, and $S_T^{\zeta}(k, \tau)$ is the scalar temperature source function. The $B$-mode fluctuation arising from primordial tensor perturbations is given by \cite{MaresukeThesis,Maresuke2010,Maresuke2011}
\be
a_B^h(\Bell) &=& \int_{0}^{\tau_0}\mathrm{d}\tau \int \frac{\mathrm{d}^3k}{(2\pi)^3} \sum_{\pm} \pm h^{\pm}(\bs{k}) e^{-i k^z D} \nonumber \\
&&  \times 2 i \frac{k^z}{k} S_P^{h}(\bs{k},\tau) (2\pi)^2\delta^{(2)}\left(\bs{k}^\parallel D-\Bell\right) \, ,
\label{eq:al_B}
\ee
where $S_P^h(k, \tau)$ is the tensor polarization source function  

After some algebra, using Eq.~\eqref{eq:flatsky}, Eq.~\eqref{eq:al_T} and Eq.~\eqref{eq:al_B} and using the primordial input spectrum Eq.~\eqref{eq:hzz_generalform} 
and applying the thin-shell approximation \cite{NonGaussianShapes} we obtain (see Appendix~\ref{app:FlatSky})
\be
B_{\Bell_1 \Bell_2 \Bell_3}^{TTB} &=& 16 \pi^2 A_s^2 \sqrt{r} f_\mathrm{NL}^{h\zeta\zeta} (\Bell_1 \times \Bell_3) \nonumber\\
&& \times \iint dk_1^z dk_2^z~\mathcal{I}\left(k_1^R,k_2^R,k_3^R\right) \nonumber\\
&& \qquad \times \Delta_{T}^{\zeta} (k_1^z, \ell_1)\Delta_{T}^{\zeta} (k_2^z, \ell_1)\Delta_{P}^h (k_3^z, \ell_3) \nonumber\\
&& \qquad \times \frac{\sqrt{2}k_3^z}{k_3^R \ell_3^2}\left[ k_1^z (\Bell_2 \cdot \Bell_3) - k_2^z (\Bell_1 \cdot \Bell_3)\right], \label{eq:BTTbispectrum2}
\ee
where the integrals over $k_1^z, k_2^z$ run from $-\infty$ to $+\infty$, $k_3^z \equiv -(k_1^z + k_2^z)$, and $k_i^R \equiv \sqrt{(k_i^z)^2 + (\ell_i/D_R)^2}$, with $D_R = \tau_0- \tau_R$, where $\tau_R$ is the conformal time at the peak of the CMB visibility function,
and, for $X = T, P$ and $s = \zeta, h$,
\be
\Delta_{X}^s (k_i^z,\ell_i) \equiv \int_0^{\tau_0} \frac{d\tau }{D^2} S_X^s(k_i^R)e^{-i k^i_z \tilde{D}}, \label{eq:Delta_Xs}
\ee
with $\tilde{D}= \tau -\tau_R$. Eq.~\eqref{eq:BTTbispectrum2} clearly takes the general form Eq.~\eqref{BTTformH}, and as a consequence has the symmetry properties discussed in Sec.~\ref{sec:envelope}.

We show  $B_{\Bell_1 \Bell_2 \Bell_3}^{TTB}$ for several slices in Fig.~\ref{fig:BTTbispectrum}. While we have chosen a particular template for the presentation here, we emphasize that in principle forecasted bounds on $f^{h\zeta\zeta}_{\rm NL}$ can be obtained for any model that predicts the coupling of two scalars and a tensor by using the appropriate $k$-dependent shape $\mathcal{I}(k_1,k_2,k_3)$.
In this paper we will use the shape of Eq.~\eqref{eq:NGshape} as an example, but will not assume a specific amplitude for $f^{h\zeta\zeta}_{\rm NL}$ unless stated otherwise. 

\begin{figure*}[t] 
  \centering
  \includegraphics[scale=.6]{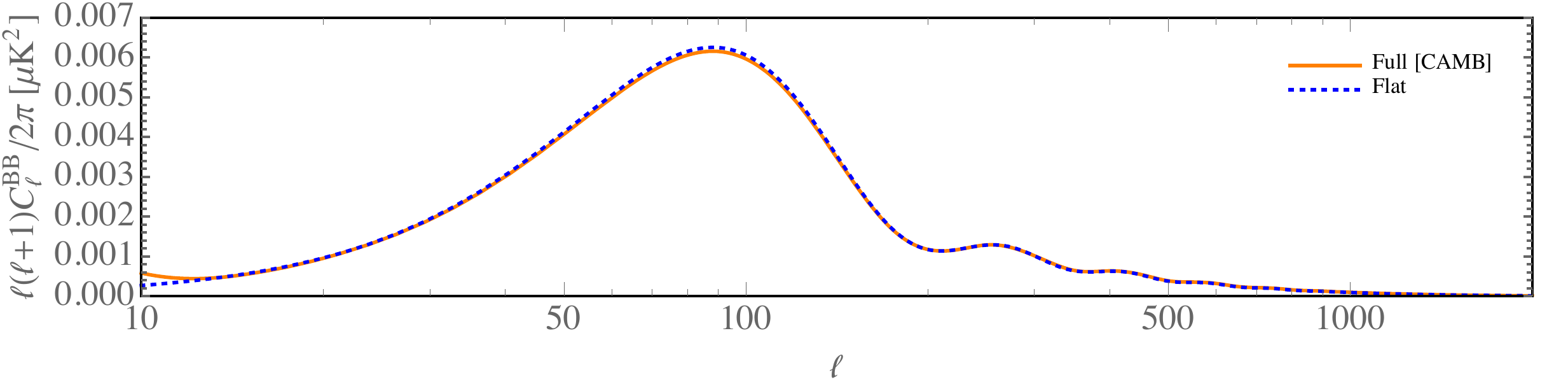}    
  \includegraphics[scale=0.293]{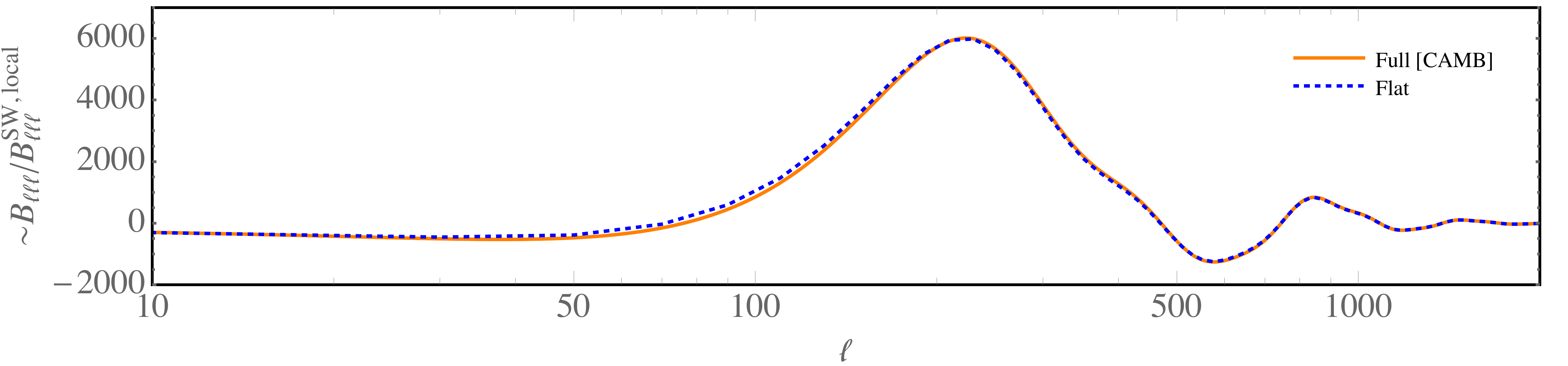}    
  \caption{Top: the B-mode power spectrum computed on the full sky and in the flat-sky, thin-shell, approximation.  The flat-sky approximation holds all the way down to $\ell \simeq 10$.  Bottom: the local-type temperature bispectrum in the full sky and flat sky (thin-shell)  for $\ell_1 = \ell_2=\ell_3$.  There are some differences on large scales, but for computing the signal-to-noise ratio we do not expect these to lead to significant deviations. Note that for purposes of presentation, the total spectrum is divided by the large scale analytical limit of the local bispectrum. \cite{FergussonBisShape2009,MoritzEtAlResonantBispectrum}. }
  \label{fig:FlatvsFullBispectrum}
\end{figure*}

\section{CMB forecasts}\label{sec:SN}

In this section, we forecast the CMB constraints on primordial tensor-scalar-scalar non-Gaussianity which can be obtained from the $\langle BTT \rangle$ bispectrum.  Our analysis does not include possible contributions to the observed $\langle BTT \rangle$ bispectrum from lensing, foregrounds, or systematic effects.  We also neglect late-time effects, analagous to those that are known to produce $\langle TTT \rangle$ correlations at the level of $f_{\mathrm{NL}}\sim 1$ even for purely Gaussian initial conditions \cite{Creminelli:2004pv, Lewis:2012tc, Pajer:2013ana, Pettinari:2014iha}.  We leave the computation of these contributions to future work.

\subsection{Qualitative considerations}

Before we compute the signal-to-noise ratio, let us make some qualitative estimates. The temperature fluctuation is mostly sourced by primordial scalar fluctuations, while the $B$-mode polarization is sourced by primordial tensor fluctuations:
\be
T &\sim& S_T ~ \zeta,\\
B &\sim& S_B ~ h.
\ee
The $\langle BTT \rangle$ and $\langle TTT \rangle$ three-point functions therefore have the following ratio:
\be
\frac{\langle BTT \rangle}{\langle TTT \rangle} \sim \frac{S_B}{S_T} \frac{\langle h \zeta \zeta \rangle}{\langle \zeta \zeta \zeta \rangle }.
\ee
We assume that the temperature fluctuation is measured up to cosmic variance, implying $\sigma_T^2 = \sigma_{T, \rm cv}^2 \sim S_T^2 \langle \zeta^2 \rangle$. In contrast, accounting for instrumental noise, the variance of the $B$-mode fluctuation is 
\be
\sigma_B^2 \sim  S_B^2 \langle h^2 \rangle \left(1 + \frac{\sigma_{B, \rm inst}^2}{\sigma_{B, \rm cv}^2}\right),
\ee
where $\sigma_{B, \rm cv}^2 \sim S_B^2 \langle h^2\rangle$ is the cosmic variance of the primordial $B$-modes and $\sigma_{B, \rm inst}^2$ is the instrumental noise (and in practice also contains residual $B$-modes from lensing and foregrounds). As a consequence, the ratio of the variance of the $\langle BTT \rangle$ estimator to that of the $\langle TTT \rangle$ estimator is 
\be
\frac{\sigma_{BTT}^2}{\sigma_{TTT}^2} &\sim& \frac{\sigma_B^2 (\sigma_T^2)^2}{(\sigma_T^2)^3} \sim \frac{\sigma_B^2 }{\sigma_T^2} \nonumber\\
&\sim&  \left(1 + \frac{\sigma_{B, \rm inst}^2}{\sigma_{B, \rm cv}^2}\right) \frac{S_B^2}{S_T^2} \times r,
\ee
where $r \equiv \langle h^2 \rangle/\langle \zeta^2 \rangle$ is the tensor-to-scalar ratio. The ratio of the signal-to-noise ratio for the $\langle BTT \rangle$ estimator to that of the $\langle TTT \rangle$ estimator is therefore
\be 
\frac{(S/N)_{BTT}^2}{(S/N)_{TTT}^2} \sim \frac1{r} \frac{\sigma_{B, \rm cv}^2}{\sigma_{B, \rm cv}^2 + \sigma_{B, \rm inst}^2} \left(\frac{\langle h \zeta \zeta \rangle}{\langle \zeta \zeta \zeta \rangle }\right)^2. \label{eq:SNR-ratio}
\ee
Now, while $\langle h^2 \rangle / \langle \zeta^2 \rangle = r \ll 1$, the ratio of the three-point functions $\langle h \zeta \zeta \rangle/ \langle \zeta \zeta \zeta \rangle$ is of order $\mathcal{O}(\varepsilon^0)$ in single-field slow-roll inflation (and \emph{not} of order $\sqrt{\varepsilon}$ as one might naively expect). We therefore see that in the limit that the instrumental noise for $B$ is subdominant to cosmic variance ($\sigma_{B, \rm inst}^2 \ll \sigma_{B, \rm cv}^2$), the expected signal-to-noise ratio of $\langle BTT \rangle$ is larger than that of $\langle TTT \rangle$ by a factor $1/\sqrt{r} \gg 1$. In the other limit  $\sigma_{B, \rm inst}^2 \gg \sigma_{B, \rm cv}^2$, this ratio saturates to a value independent of $r$ (since $\sigma_{B, \rm cv}^2 \propto r$).

Similarly, the advantage of using $\langle BTT\rangle$ over $\langle TTT\rangle$ to constrain $\langle h\zeta\zeta\rangle$ is immediately apparent from Eq.~\eqref{eq:SNR-ratio} by replacing $\langle h \zeta \zeta \rangle/ \langle \zeta \zeta \zeta \rangle \rightarrow 1$, showing that for sufficiently low values of $r$, $\langle BTT\rangle$ will always provide a better constraint than $\langle TTT \rangle$ on $\langle h\zeta\zeta \rangle$ {\it independent of the model}.

Therefore, we see that while lower noise measurements of temperature fluctuations will not lead to significant further improvement of the measurement of primordial non-Gaussianity from $\langle TTT \rangle$, there is a great deal of room for improvement with $\langle BTT \rangle$ as a probe of primordial physics. 

\subsection{Quantitative calculation}

In the flat-sky approximation the signal-to-noise ratio is given by the following integral over multipoles (rather than a discrete sum in the full-sky case) \cite{HU2000SN}: 
\be
\left(\frac{S}{N}\right)^2 = \frac{f_{\rm sky}}{4\pi^3} \int d^2\Bell_2 \int d^2 \Bell_3  \frac{\left(B_{\left(-\Bell_2 - \Bell_3\right),  \Bell_2, \Bell_3}^{TTB}\right)^2}{\mathcal{C}^{TT}_{\ell_1} \mathcal{C}^{TT}_{\ell_2} \mathcal{C}^{BB}_{\ell_3} }, 
\label{eq:2DSN}
\ee
where $\mathcal{C}_{\ell} = C_{\ell} + N_{\ell}$ with $C_{\ell}$ is the angular power spectrum, $N_{\ell}$ the noise, and we integrated out the $\Bell_1$ direction. 
\hspace*{-3cm}
\begin{figure}[t] 
  \includegraphics[scale=.34]{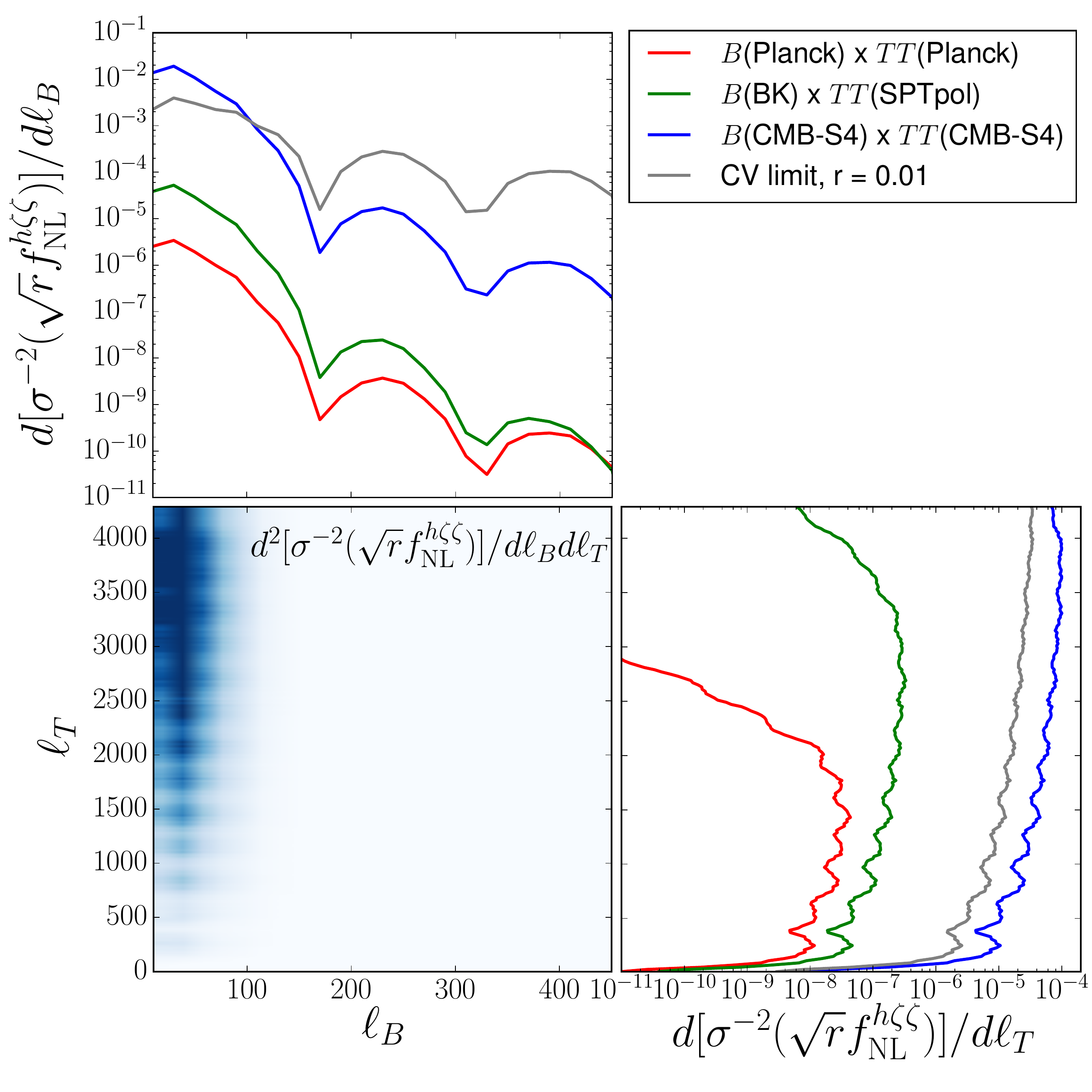}    
  \caption{Density plot showing the CMB modes to be measured in order to obtain a detection of the \BTT\ bispectrum, showing the contributions to the inverse noise or Fisher matrix element.  The color map on the bottom left, which uses a linear scale, shows the contribution  as a function of the $B$ multipole and one of the two $T$ multipoles for the CMB-S4 case.  The other panels show the collapsed one-dimensional distributions for CMB-S4 as well as three other cases.  The signal is concentrated in slices with small $\ell_B$ and a wide range of values of $\ell_T$.  We show constraints from several experiments, assuming no sample variance in the $B$ modes, as well as with the cosmic variance limit for $r = 0.01$.  }
  \label{fig:SNBTTdensity}
\end{figure}

To compute the signal-to-noise ratio, we modified CAMB \cite{lewis99,lewis13} and CLASS \cite{CLASS} to extract the scalar and polarization source functions. We then computed the bispectrum using Eq.~\eqref{eq:BTTbispectrum2} and the signal-to-noise ratio using Eq.~\eqref{eq:2DSN}. For consistency we also used the flat-sky $C_{\ell}$ for our variance estimate. We compared our $C_{\ell}$ to the full-sky results and found good agreement for $\ell \gtrsim 10$ as can be seen in the top panel of Fig.~\ref{fig:FlatvsFullBispectrum}. We also show the diagonal of the $\langle TTT \rangle$ flat sky bispectrum versus the full-sky version in the bottom panel of Fig~\ref{fig:FlatvsFullBispectrum}. Unlike Ref.~\cite{FergussonBisShape2009}, we  find good agreement all the way down to $\ell \sim 10-20$. For purposes of computing the signal-to-noise ratio of the $\langle BTT \rangle$ bispectrum, the small amplitude differences are not a concern. That being said, there is potentially a significant contribution to the signal-to-noise ratio in the lowest-$\ell$ $B$-modes, given the presence of the reionization bump at low $\ell$ which is not included due to the approximations we have made here.  We will leave the full-sky computation including these modes for a future study. 

All results are obtained using a Planck fiducial cosmology, $k_{\rm max} = 0.64$ Mpc$^{-1}$ for scalars and  $k_{\rm max} = 0.17$ Mpc$^{-1}$ for tensors.  
We performed calculations to maximal multipoles of $\ell_{T,{\rm max}} = 4500$ for the $T$ modes and to $\ell_{{B,\rm max}} = 500$ for the $B$ modes.  On small scales in the temperature, foreground fluctuations  such as the emission from dusty, star-forming galaxies and the thermal Sunyaev-Zel'dovich effect will reduce the effective maximal temperature multipole for primordial studies to about 3000--4000.  Given multifrequency data, these could in principle be removed, leaving the kinetic Sunyaev-Zel'dovich effect as the dominant foreground on small scales.  Depending on the amplitude of this signal, this could make the maximal multipole as large as  $\ell_{T,\rm max} \sim 4500$ \cite{georgeSPT}.  

\begin{figure}[t] 
  \centering
  \includegraphics[scale=.55]{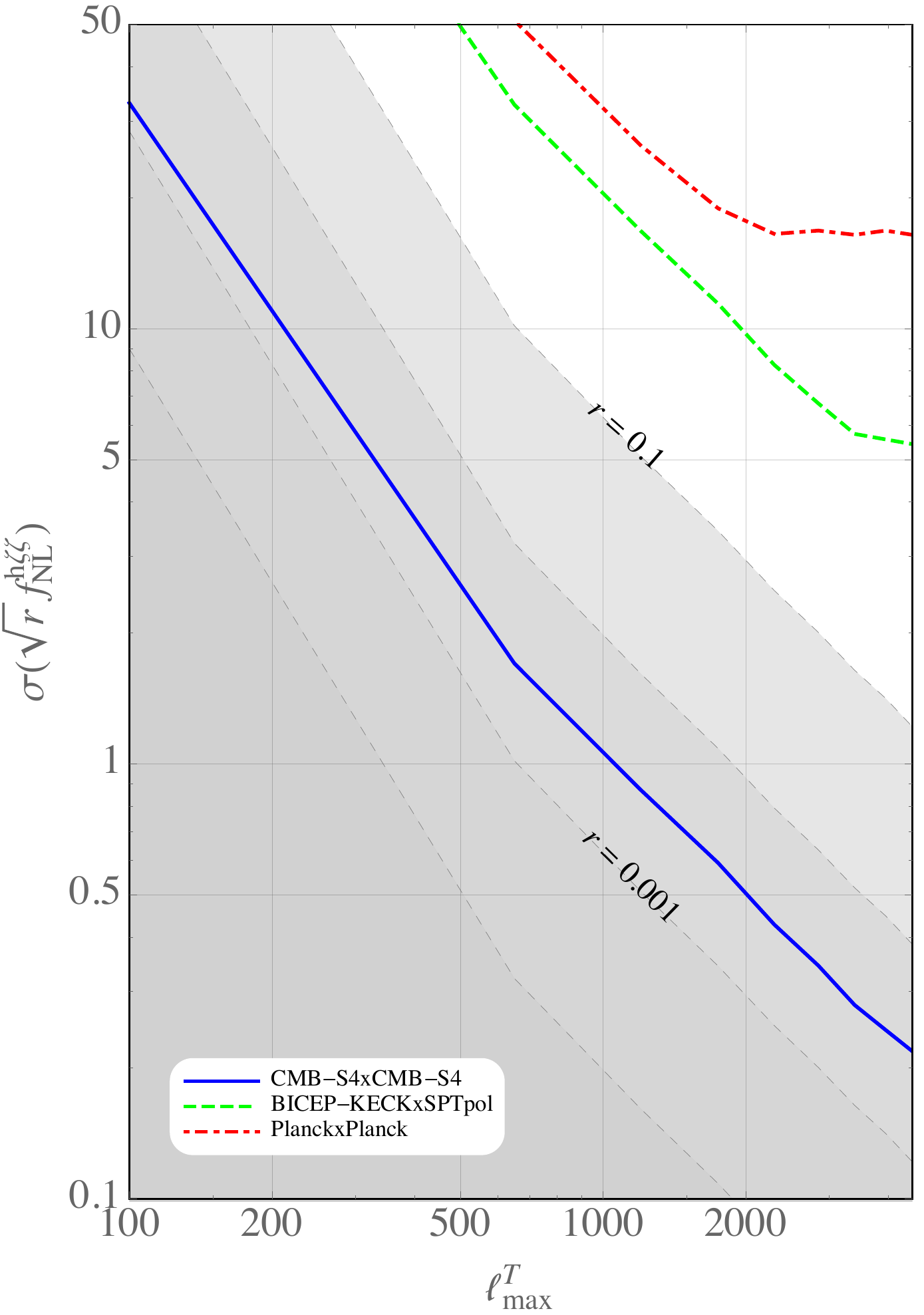}    
  \caption{Forecasts for $\sigma (\sqrt{r} f_{\rm NL}^{h\zeta\zeta})$ for various CMB experiments as a function of $\ell_{\rm max}$. The colored lines present constraints when cosmic variance is negligible. The figure shows that cosmic variance would be subdominant for current and near future experiments if $r=0.01$. For an experiment like CMB-Stage IV the total variance would be dominated by cosmic variance and not by instrumental noise unless $r \lesssim 0.001$ (with $\ell^T_{\rm max} = 4500$).}
  \label{fig:DeltaFNL_BTT}
\end{figure}

The bispectra are computed on a grid with $\Delta \ell = 1$ for $\ell \leq 100$, $\Delta \ell = 10$ for $100 \leq \ell \leq 500$ and $\Delta \ell = 20$ for all values above $\ell = 500$. We use linear interpolation  to obtain the curves shown in Fig.~\ref{fig:SNBTTdensity} and \ref{fig:DeltaFNL_BTT}.


\begin{table*}
  \begin{center}
    \begin{threeparttable}

      \begin{tabular}{|  l  c  c c || l c c c || c |}
        \hline
        $B$ expt.  & $B$ noise ($\mu$K-') & $B$ beam (') & $\ell_b$ & $T$ expt. & $T$ noise ($\mu$K-') & $T$ beam (') & $\ell_b$ & area  (sq. deg.)\\
        \hline
        Planck     &   60         &  5       & 1600      & Planck & 30 & 5 & 1600 & 33,000 \\
        BICEP/Keck &   3          & 60       & 130       & SPTpol & 5  & 1 & 8100 & 625 \\
        CMB-S4     &   $\sqrt{2}$ &  1       & 8100      & CMB-S4 & 1  & 1 & 8100 & 33,000\\
        \hline
      \end{tabular}
      
    \end{threeparttable}
  \end{center}
  \caption{Assumed experimental parameters for forecasts.  Beamsizes in arcmin are quoted as FWHM,  related to $\sigma_b$ in  Eq. \ref{eq:noisepower} by a factor of $2 \sqrt{2 \ln 2}$.  For reference we also show $\ell_b \equiv 1/\sigma_b$.}
  \label{tab:expts}
\end{table*}

We assume that the noise power spectra in the $T$ and $B$ measurements are given by 
\begin{equation}
N_{\ell}^{YY} = (w_0^Y)^2 \exp(\ell^2 \sigma_b^2),
\label{eq:noisepower}
\end{equation}
for $Y \in \{T, B\}$, and where the noise levels $w_0$ and beamsizes $\sigma_b$ are given in Table~\ref{tab:expts}.  

\begin{figure}[t] 
  \includegraphics[scale=.34]{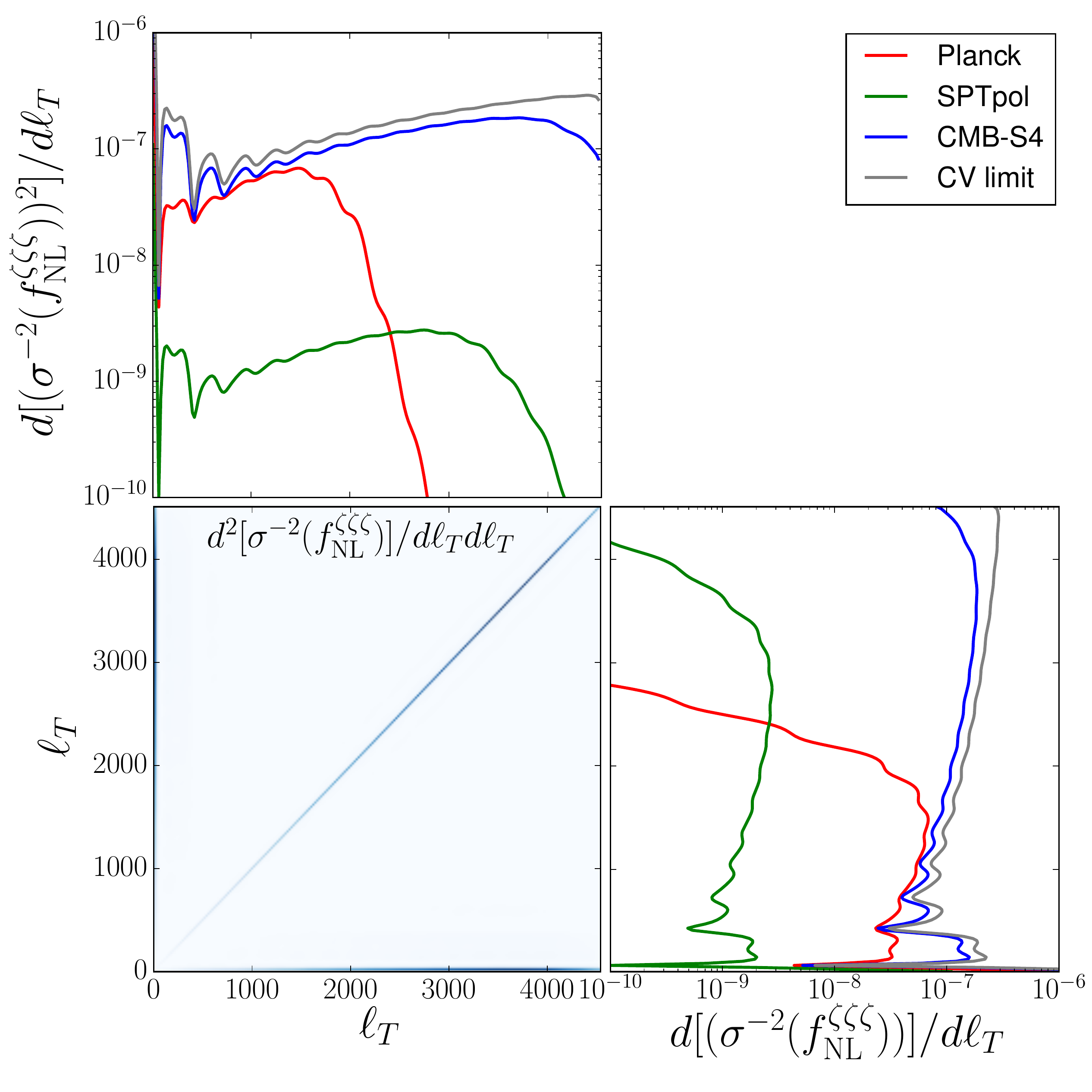}    
  \caption{Same as \ref{fig:SNBTTdensity}, but for the local-type TTT bispectrum, sourced by scalars.  The signal is concentrated at very low $\ell_T$ and along the diagonal, where both $\ell_T$  are equal. }
  \label{fig:SNTTTdensity}
\end{figure}

In order to better understand which modes contribute to the signal,  in Fig.~\ref{fig:SNBTTdensity} we show the inverse-variance density, defined using the free parameters in our parameterization, as ${d^2[\sigma^{-2}({\sqrt{r} f_\mathrm{NL}^{h\zeta\zeta} })]}/{d\ell_T d\ell_B}$.
This is equivalent to the signal-to-noise ratio, Eq.~\ref{eq:2DSN}, for a model with ${\sqrt{r} f_\mathrm{NL}^{h\zeta\zeta} } = 1$.  As expected, the signal is concentrated on slices with $\ell_B \ll \ell_T$, and $T$ modes contribute down to very small scales ($\ell_T \simeq$ several thousand).  
For comparison, we show a similar plot for the $\langle TTT \rangle$ bispectrum resulting from local-type primordial non-Gaussianity in Fig.~\ref{fig:SNTTTdensity}.  

\begin{figure}[t] 
  \centering
  \includegraphics[scale=.546 ]{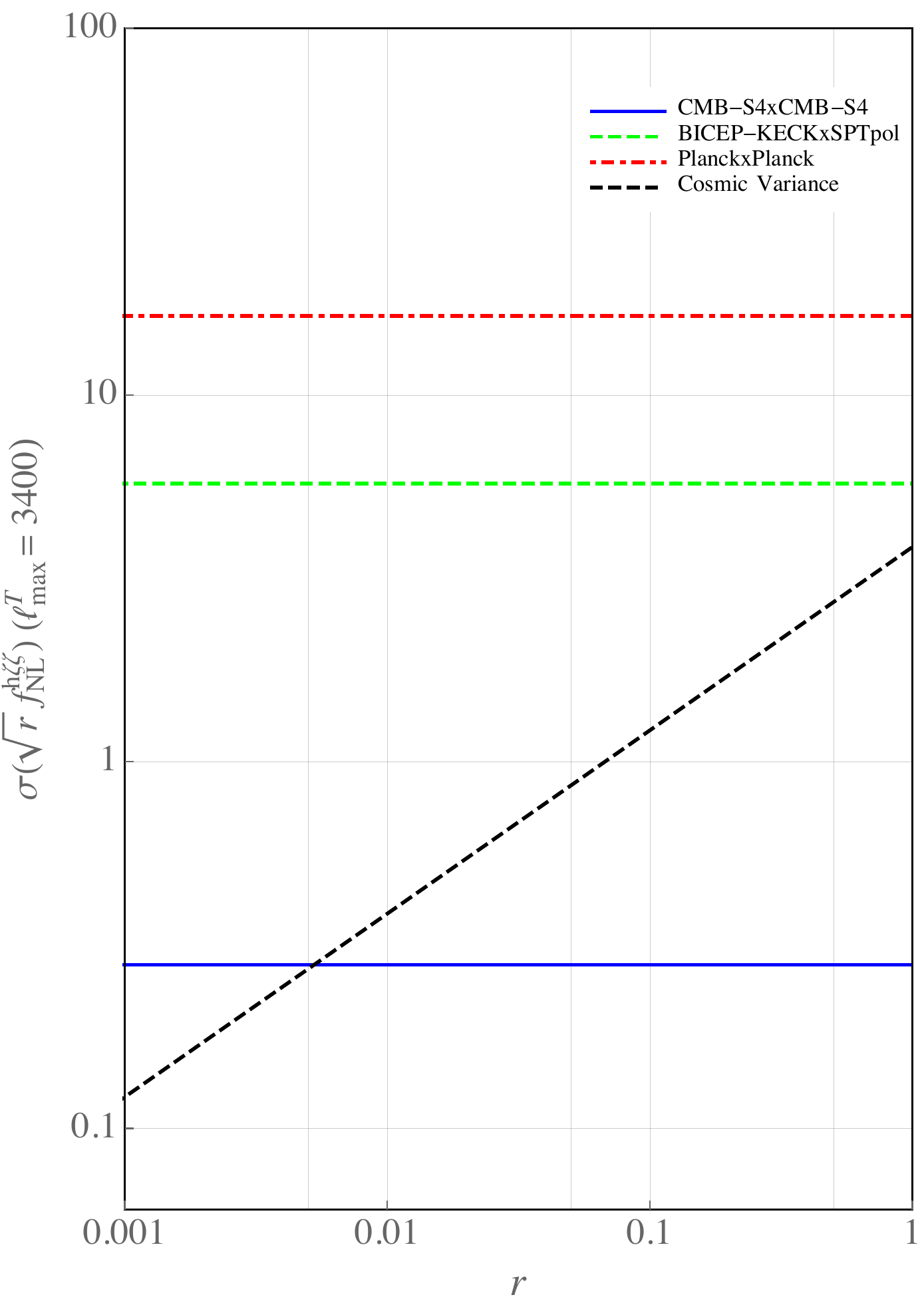}    
  \caption{Forecasts for $\sigma(\sqrt{r} f_{\rm NL}^{h\zeta\zeta})$ for various CMB experiments. This figure illustrates that current experiments are all noise dominated for allowed values of $r$. CMB-Stage IV is comic variance dominated unless $r \lesssim 0.005$ (with $\ell^T_{\rm max} = 3400$).  Cosmic variance limit can only be reduced if we consider more modes, i.e. by increasing $\ell_{\rm max}^T$).}
  \label{fig:fnlalt}
\end{figure}

In Fig.~\ref{fig:DeltaFNL_BTT} and Fig.~\ref{fig:fnlalt} we show the forecasts for an analysis with publicly-available data,  $B$(Planck)$\times TT$(Planck); currently-taken data,  $B$(BICEP/Keck)$\times TT$(SPTpol);  and futuristic data, $B$(CMB-S4)$\times TT$(CMB-S4).  Although not shown, a measurement from $B$(SPIDER)$\times TT$(SPT)  \cite{SPIDER} would be similar to that from  BICEP/Keck and SPTpol: the noise levels are lower, but that is offset by the relatively small sky coverage of BICEP/Keck-SPTpol. We consider two distinct scenarios, represented by the limiting cases of Eq.~\eqref{eq:SNR-ratio}: first, the case that there is no signal from primordial tensors where we set ${C}_{\ell}^{BB} = N_{\ell}^{BB}$; second, the  $B$-mode cosmic variance limit for various values of the tensor-to-scalar ratio $r$.   All bounds are shown as  functions of $\ell_{T{\rm max}}$;  Fig.~\ref{fig:SNBTTdensity} shows that multipoles with $\ell_B \gtrsim 100$ hardly contribute to the final bounds.  

In single-field slow-roll inflation, it is predicted that $f_{\rm NL}^{h\zeta\zeta}=\sqrt{r}/16$, and so in the event that cosmological $B$-modes are detected, this will provide a consistency check on the model.  As can be seen from Fig.~\ref{fig:fnlalt}, sample variance prevents us from detecting $f_{\rm NL}^{h\zeta\zeta}$ from $\langle BTT \rangle$ if single-field slow-roll inflation is the source of the fluctuations we observe.  A detection of $f_{\rm NL}^{h\zeta\zeta}$ would therefore imply that single-field slow-roll inflation is not solely responsible for the observed fluctuations.  


With an experiment like CMB-Stage IV \cite{Abazajian:2013oma} we anticipate that we can constrain $\sqrt{r} f_{\rm NL}^{h\zeta\zeta} \sim \mathcal{O}(0.1) f_{\rm sky}^{-1/2}$.  It is remarkable that the potential constraint on primordial non-Gaussianity using $\langle BTT \rangle$, which, given a similar shape and normalization of the tensor-scalar-scalar bispectrum and the scalar-scalar-scalar bispectrum, lies an order of magnitude below the optimal CMB constraint on the local-type scalar non-Gaussianity $f_{\rm NL}^{\zeta\zeta\zeta}$. Furthermore, it was shown in Ref.~\cite{MaresukeThesis} that the $\langle TTT \rangle$ bispectrum could provide constraints on $\sqrt{r} f_{\rm NL}^{h\zeta\zeta}\sim \mathcal{O}(10) f_{\rm sky}^{-1/2}$. The forecasts presented in this paper show that using $\langle BTT \rangle$ has the potential to improve that constraint by nearly two orders of magnitude. In order to fully exploit the power of $\langle BTT \rangle$ one should consider more general models of the early Universe that could potentially violate this bound. We will leave this to future work.

\section{Discussion and conclusion}\label{sec:Conclusions}

We have explored the potential of the $\langle BTT \rangle$ bispectrum as probe of the early universe. The odd intrinsic parity of $B$-modes gives this bispectrum some properties which differ from that of the $\langle TTT \rangle$ bispectrum, but both are generically non-vanishing in a parity-conserving universe, and are sourced by primordial bispectra which are predicted to be of the same order in slow-roll parameters in single-field slow-roll inflation.

One advantage of the $\langle BTT \rangle$ bispectrum is that the signal suffers less from cosmic variance than its $\langle TTT \rangle$ counterpart for constraining the tensor-scalar-scalar bispectrum.  Our analysis shows that with this observable it should be possible to constrain the level of non-Gaussianity to $\sigma(\sqrt{r}f_{\rm NL}^{h\zeta\zeta}) \sim \mathcal{O}(0.1) f_{\rm sky}^{-1/2}$. For comparison, the CMB can only measure the local-type scalar bispectrum down to $\sigma (f_{\rm NL}^{\zeta \zeta \zeta}) \sim \mathcal{O}(1)$ \cite{FnlForecast2007,PlanckNonGaussianity2015} and with upcoming large-scale-structure surveys aimed at constraining non-Gaussianity using galaxies, $\sigma ( f_{\rm NL}^{\zeta \zeta \zeta}) \sim 0.2$ \cite{LSSnonGaussianity2014}.

Our analysis was done in the flat-sky limit, valid for multipoles $\ell \gtrsim 10$, which suffices for ground-based detectors. It will be important to extend this analysis to the full sky since there is potentially significant signal in the lowest $\ell$ modes, given the presence of the relatively large $B$ fluctuation from scattering at reionization. This signal  can most likely only be mapped out by a dedicated satellite, such as the proposed LiteBIRD \cite{LiteBIRD}, PIXIE \cite{PIXIE} or COrE \cite{CORE} experiments. 
  
We focused in this work on a template for the $\langle BTT \rangle$ bispectrum motivated by single-field slow-roll inflation, which maximizes when large-scale $B$ modes are correlated with small-scale $T$ modes.  This shape has some experimental advantages, since a search for such a bispectrum could be performed, for instance, by cross-correlating a map of $B$-modes on large scales from a current or upcoming ground-based CMB experiment with small-scale $T$ fluctuations, such as those measured with the Planck satellite.  On the other hand, it would be useful to consider other shapes for the $\langle BTT \rangle$ bispectrum, perhaps motivated by specific early-universe models.

In this work we have not considered contamination from dust, systematics,  or lensing.   Dust is a well-known contaminant in estimates of the \BB\ power spectrum on large scales at the frequencies probed with ground-based experiments.   Estimates of the \BTT\ bispectrum will in principle be sensitive to correlations between large-scale dust polarization and small-scale dust intensity; while this may be less of an issue than in the power spectrum measure, this needs to be investigated in future work.  Similarly, instrumental systematics which affect the measure of $B$ on large scales should be decoupled from those that affect $T$ on small scales, making this analysis less sensitive to systematics than the  \BB\ power spectrum.  Finally, lensing converts $E$-mode polarization to $B$-mode polarization.  As with measurements of the \BB\ power spectrum, delensing to reduce effective noise from lensing needs to be performed when measuring the \BTT\ bispectrum.  In a universe with primordial gravitational waves, the \BTT\ bispectrum will also contain a non-primordial signal on very large scales arising from correlations between $B$-modes from Thomson scattering after reionization and the curl mode of CMB lensing, which affects pairs of temperature modes \cite{cooray2005}.  This is analogous to the  scattered $E$-mode--lensing correlation induced by scalars in estimates of the $\langle ETT \rangle$ bispectrum   \cite{lewis2011}.

We have focused on the $\langle BTT \rangle$ correlation function. However, other combinations sensitive to the coupling between scalars and tensors will add to the total signal-to-noise ratio. In particular, $\langle BEE \rangle$ and $\langle BTE \rangle$ are expected to have similar constraining power. 
In addition, similar to the use of $E$-modes for the scalar bispectrum \cite{FnlForecast2007}, the $B$ and $E$ modes are projected through functions that have different nulls, which improves the mapping from the primordial space.
In summary, the $\langle BTT \rangle$ bispectrum and other non-Gaussian correlations involving $B$-modes open up a new window into the early Universe.  Ongoing and future CMB experiments will naturally make observations which allow us to carry out searches for and place non-trivial constraints on primordial tensor non-Gaussianity.  While more theoretical work remains to discover the full value of $B$-mode non-Gaussianity, this new set of observables has the potential to be a very rich set of tools for probing primordial physics.

\section*{Acknowledgments} 

We thank Raphael Flauger, Marc Kamionkowski, Antony Lewis, Suvodip Mukherjee, Moritz M\"unchmeyer, Ue-Li Pen, Sasha Rahlin, Fabian Schmidt, David Spergel, and Aaron Zimmerman for useful discussions. We would especially like to thank Enrico Pajer for providing us detailed comments and feedback on the draft. PDM would like to thank the hospitality of Marc Kamionkowski and Johns Hopkins University where part of this work was completed. YAH is supported by the John Templeton Foundation.

\vspace{1cm}
\begin{widetext}
  \appendix
    \section{The flat-sky $\langle BTT \rangle$ bispectrum}\label{app:FlatSky}

In this appendix, we work out the form of the three-point function $\langle BTT \rangle$ in the flat-sky approximation. Our derivation follows closely that of the Appendix of Ref.~\cite{NonGaussianShapes}. 

We start by rewriting $D = D_R + \tilde{D}$, where $D_R \equiv \tau_0 - \tau_R$ and $\tilde{D} \equiv \tau_R - \tau$, in the exponential factors of Eqs.~\eqref{eq:al_T} and \eqref{eq:al_B}. The Dirac delta function in the primordial three-point function ensures that the $D_R$ dependence drops of the exponential factors. Using Eqs.~\eqref{eq:al_T} and \eqref{eq:al_B}, we then get the following three-point function
 
 \begin{align}
      \left\langle a_T^\zeta(\Bell_1) a_T^\zeta(\Bell_2) a_B^h(\Bell_3) \right\rangle &= \prod_{n=1}^3 \left[\int_{0}^{\tau_0} \mathrm{d} \tau_n \int \frac{\mathrm{d}^3k_{n}}{(2\pi)^3} e^{-ik_n^z \tilde{D}_n} (2\pi)^2\delta^{(2)}\left(D_n \bs{k}_n^\parallel+\Bell_n\right) \right] \non
      &\times S_T^\zeta(k_1,\tau_1) S_T^\zeta(k_2,\tau_2) \frac{2 i k_3^z}{k_3} S_{P}^h(k_{3},\tau_3) \sum_{\pm} \pm \left\langle \zeta\left(\bs{k}_1 \right) \zeta\left(\bs{k}_2 \right) h^{\pm}\left(\bs{k}_3 \right) \right\rangle. 
      \end{align}
We now define 
\beq
\frac{2 i k_3^z}{k_3} \sum_{\pm} \pm \left\langle \zeta\left(\bs{k}_1 \right) \zeta\left(\bs{k}_2 \right) h^{\pm}\left(\bs{k}_3 \right) \right\rangle \equiv (2 \pi)^3 F(\bs{k}_1, \bs{k}_2, \bs{k}_3) \delta^{(3)}\left(\sum_n\bs{k}_n\right). \label{eq:F_kernel}
\eeq
Assuming the kernel $F$ does not vary significantly across the width of the last-scattering surface (i.e. specifically, that it does not change much as $\bs{k}_i$ varies by a fractional amount $\Delta \tau_R/\tau_R \sim 10^{-2}$ \cite{NonGaussianShapes}), we can take it out of the time integrals, by setting $\bs{k}_n^{||} \approx \bs{\ell}_n/D_R$ inside $F$. Using $\bs{k}_n^{||} = \bs{\ell}_n/D_n$, the Dirac function in Eq.~\eqref{eq:F_kernel} simplifies to
\beq
 \delta^{(3)}\left(\sum_n\bs{k}_n\right) = \delta^{(1)}\left(\sum_n\bs{k}_n^z\right) \times D_R^2 \delta^{(2)} \left( \sum_n \bs{\ell}_n + \sum_n \frac{\tilde{D}_n}{D_R} \bs{\ell}_n \right) \approx D_R^2\delta^{(1)}\left(\sum_n\bs{k}_n^z\right) \delta^{(2)} \left( \sum_n \bs{\ell}_n \right).
\eeq
Because the last equality is only approximate, the bispectrum does not exactly vanish for modes that do not form a closed triangle; however, it is exponentially suppressed on these configurations \cite{Boubekeur:2009uk}.

These approximations allow to factorize the expression for the bispectrum. Defining $\Delta_{X}^s(k_i^z, \ell_i)$ as in Eq.~\eqref{eq:Delta_Xs},
we arrive at
\be
 \left\langle a_T^\zeta(\Bell_1) a_T^\zeta(\Bell_2) a_B^h(\Bell_3) \right\rangle \approx  \delta^{(2)}(\bs{\ell}_1 + \bs{\ell}_2 + \Bell_3) \iint d k_1^z d k_2^z  \Delta_T^\zeta(k_1^z, \ell_1)\Delta_T^\zeta(k_2^z, \ell_2)\Delta_P^h(k_3^z, \ell_3) 
 F(\bs{k}_1, \bs{k}_2, \bs{k}_3),\label{eq:flat-sky-general}
 \ee
where the integrand is to be evaluated at $k_3^z = - (k_1^z + k_2^z)$ and $\bs{k}_n^{||} = \bs{\ell}_n/D_R$ inside $F$.

The last step is to derive an explicit expression for $F$.  Using Eq.~\eqref{eq:hzz_generalform} and our definition \eqref{eq:F_kernel} we get
\beq
F(\bs{k}_1, \bs{k}_2, \bs{k}_3) = 16 \pi^4 A_s^2 f_{\rm NL}^{h \zeta \zeta} \mathcal{I}(k_1, k_2, k_3)  2^{3/2}\frac{k_3^z}{k_3}  G(\bs{k}_1, \bs{k}_2, \bs{k}_3) , \label{eq:Fki}
\eeq
where we have defined
\beq
G(\bs{k}_1, \bs{k}_2, \bs{k}_3) \equiv D_R^2 \frac{i}{\sqrt{2}}   \sum_{\pm}\pm e_{ab}^{\mp}(\bs{k}_3)k_1^ak_2^b. 
\eeq
Using Eq.~\eqref{eq:e_ab}, we have, in a basis whose third axis is along $\bs{k}_3$,
\beq
\sum_{\pm} \pm e_{ab}^{\mp}(\bs{k}_3) = - i \sqrt{2} \begin{pmatrix}
      0       & 1 & 0  \\
      1     & 0 & 0 \\
      0       & 0 & 0 
    \end{pmatrix}.
\eeq
We can find the polarization tensor for an arbitrary direction by performing a standard rotation on each axis of the polarization tensor.  For $\hat{n} = (\sin\theta\cos\phi,\sin\theta\sin\phi,\cos\theta)$, the standard rotation matrix $S(\hat{n})$ is given by \cite{weinberg2008cosmology}
    \be
    S_{ab}(\hat{n}) \equiv
    \begin{pmatrix}
      \cos\theta\cos\phi & -\sin\phi & \sin\theta\cos\phi \\
      \cos\theta\sin\phi & \cos\phi & \sin\theta\sin\phi \\
      -\sin\theta & 0 & \cos\theta
    \end{pmatrix} \, .
    \ee
    The polarization sum for $\bs{k}_3 = k_3 (\sin\theta\cos\phi,\sin\theta\sin\phi,\cos\theta)$ in an arbitrary direction is such that
    \be
\frac1{D_R^2} G(\bs{k}_1, \bs{k}_2, \bs{k}_3)  =&  \sin\theta\left[-\left(k_1^zk_{2}^y+k_{1}^y k_2^z\right)\cos\phi + \left(k_1^zk_{2}^x+k_{1}^x k_2^z\right)\sin\phi\right] \non
      &+\cos\theta\left[\left(k_{1}^y k_{2}^x+k_{1}^x k_{2}^y\right)\cos(2\phi)+\left(-k_{1}^x k_{2}^x+k_{1}^y k_{2}^y\right)\sin(2\phi)\right] \, .
    \ee
    Due to rotational invariance, we can, without loss of generality, choose $\bs{k}_3$ to have vanishing $y$-component. Once we compute the bispectrum for this choice of $\bs{k}_3$, all bispectra with general $\bs{k}_3$ can be obtained by rotation about the $z$-axis.  For $\bs{k}_3$ in the $xz$-plane, the polarization sum gives
    \be
 \frac1{D_R^2} G(\bs{k}_1, \bs{k}_2, \bs{k}_3)
      = (k_{1}^yk_{2}^x+k_{1}^x k_{2}^y)\frac{k_3^z}{k_3} - (k_{1}^y k_2^z+k_1^zk_{2}^y)\frac{k_{3}^x}{k_3} = \frac{k_1^y}{k_3} (k_2^x k_3^z - k_2^z k_3^x) + (1 \leftrightarrow 2).
\ee
Using the triangle condition, and substituting $\bs{k}_i^{||} = (k_i^x, k_i^y)$ by $\bs{\ell}_i/D_R$, we rewrite this expression as
\beq
G(\bs{k}_1, \bs{k}_2, \bs{k}_3) =  \frac{\ell_1^y}{k_3}(k_2^z \ell_1^x  - k_1^z \ell_2^x) + (1 \leftrightarrow 2) = 2\frac{\ell_1^y}{k_3}(k_2^z \ell_1^x  - k_1^z \ell_2^x), 
\eeq 
where the last equality arises from the fact that we chose $\bs{k}_3$ in the $xz$-plane, implying that $\ell_3^y = 0 = \ell_1^y + \ell_2^y$. 

Finally, we may rewrite $\ell_1^y = \hat{\ell}_3 \times \Bell_1$ and $\ell_1^x = \hat{\ell}_3 \cdot \Bell_1$, arriving at
\beq
G(\bs{k}_1, \bs{k}_2, \bs{k}_3) =  \frac{2}{ k_3 \ell_3^2} (\Bell_1 \times \Bell_3) \left[k_1^z (\Bell_2 \cdot \Bell_3)  - k_2^z (\Bell_1 \cdot \Bell_3)\right],
\eeq
an expression which is symmetric under exchange of $(1 \leftrightarrow 2)$ since $\Bell_2 \times \Bell_3 = - \Bell_1 \times \Bell_3$. The scalar products can be expressed in terms of magnitudes through $2\Bell_1 \cdot \Bell_3 = \ell_2^2 - \ell_1^2 - \ell_3^2$.

Inserting this expression into Eq.~\eqref{eq:Fki} and then into Eq.~\eqref{eq:flat-sky-general} gives the final expression for the flat-sky bispectrum, Eq.~\eqref{eq:BTTbispectrum2}. Using the property $\Delta_X^s(-k_i^z) = \Delta_X^s(k_i^z)^*$, and the fact that $F(\bs{k}_1, \bs{k}_2, \bs{k}_3) \propto k_3^z k_1^z , k_3^z k_2^z$, one can easily show that the three-point function \eqref{eq:flat-sky-general} is real, as it should. Finally, we also see that it has the same form as that derived in Eq.~\eqref{BTTformH} from symmetry considerations.

\end{widetext}
\bibliography{BTTbib}

\end{document}